\documentclass[aps,pre,reprint,groupedaddress]{revtex4-1}

\usepackage{amsmath,amssymb}
\usepackage{graphicx}

\begin{document}


\title{Frustration in vicinity of transition point of Ising spin glasses}


\author{Ryoji Miyazaki}
\affiliation{Department of Physics, Tokyo Institute of Technology, Oh-okayama, Meguro-ku, Tokyo 152-8551, Japan}


\date{\today}

\begin{abstract}
We conjecture the existence of a relationship between frustration and the transition point at zero temperature of Ising spin glasses.
The relation reveals that, in several Ising spin glass models, the concentration of ferromagnetic bonds is close to the critical concentration at zero temperature when the output of a function about frustration is equal to unity.
The function is the derivative of the average number of frustrated plaquettes with respect to the average number of antiferromagnetic bonds.
This relation is conjectured in Ising spin glasses with binary couplings on two-dimensional lattices, hierarchical lattices, and three-body Ising spin glasses with binary couplings on two-dimensional lattices.
In addition, the same argument in the Sherrington-Kirkpatrick model yields a point that is identical to the replica-symmetric solution of the transition point at zero temperature.
\end{abstract}


\maketitle

\section{Introduction}

Phase transition is one of the fundamental issues in physics.
Most of the successes on the topics in the field of spin glasses~\cite{Mezard, Nishimori} have been limited to the mean-field models~\cite{Sherrington}.
Unveiling the structure of phase diagrams in finite-dimensional spin glasses remains one of the most challenging problems.
In particular, although the phase transition to the spin-glass phase is a characteristic phenomenon in spin glasses, we have not had a clear understanding of its nature in finite dimensions.
Moreover, the determination of the order-disorder phase boundaries in spin glasses has a practical significance beyond pure physicists' interest, since the boundaries in spin glasses correspond to the accuracy threshold in topological quantum error-correcting codes~\cite{Dennis}.

\begin{figure}[b]
  \begin{center}
  \includegraphics[width=4cm,clip]{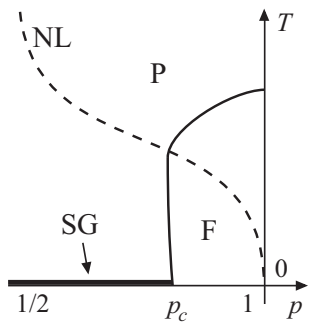}
  \caption{Phase diagram of the $\pm J$ Ising model in two dimensions.
The vertical and horizontal axes express the temperature and the concentration of ferromagnetic bonds, respectively.
The symbols P, F, and SG denote the paramagnetic, ferromagnetic, and spin-glass phases, respectively.
The Nishimori line (NL) is drawn as a dashed line.
The critical concentration at zero temperature is denoted by $p_{c}$.}
  \label{fig:pd}
  \end{center}
\end{figure}

Concerning the phase diagram, there has been a proposition based on the property of the Nishimori line (NL)~\cite{Nishimori2, Nishimori}, which is a special line across the phase diagram depicted in Fig.~\ref{fig:pd}, 
that the phase transition between the ferromagnetic and non-ferromagnetic phases at lower temperature than the NL is induced by a geometric nature~\cite{Nishimori, Nishimori3}.The form of the entropy of frustration distribution is identical to the form of the free energy on the NL.
Frustration is a geometric quantity to be defined later.
Since phase transitions are usually signaled by singularities in the free energy, the entropy of frustration distribution also has a singularity at the transition point on the NL.
This is the origin of the geometry-induced phase transition.
However, this proposition is a conjecture since the singularities in the free energy and the entropy of frustration distribution do not necessarily coincide in regions other than the NL.
The case of the NL, nevertheless, suggests the potential of frustration in the phase transitions in spin glasses.

In order to illustrate the concept of frustration, we consider the $\pm J$ Ising model on a square lattice governed by the Hamiltonian 
\begin{equation}
H = - \sum_{<ij>} J_{ij} S_{i} S_{j}, \label{eq:pmJ}
\end{equation}
where $J_{ij}$ takes either $J (>0)$ with probability $p$ or $-J$ with probability $1-p$, and $S_{i}$ is an Ising variable ($S_{i} = \pm 1$).
The summation runs over nearest neighbors.
If the number of negative (antiferromagnetic) coupling constants is odd in a loop composed of bonds, there is no spin configuration, permitting all bonds to be in the lower-energy state ($-J_{ij} S_{i} S_{j} = -J$).
The product of coupling constants $\prod_{<ij> \in c} J_{ij}$ over an arbitrary loop $c$ is called frustration~\cite{Toulouse, Nishimori}.
If the frustration of a loop has a negative value, the loop is regarded as a ``frustrated loop".

Motivated by the argument about the geometry-induced phase transition, we aim to detect phase transitions only with consideration of frustration in the present study.
More specifically, we focus our attention on the number of frustrated plaquettes in the Ising spin glass models.
Note that a ``plaquette'' means an elementary loop composed of bonds on the lattice, which cannot be divided into any other loops.
We observe the change in the average number of frustrated plaquettes as the average number of
antiferromagnetic bonds is increased.

The present paper is organized as follows.
In Sect.~2 we consider the $\pm J$ Ising model on the two-dimensional lattices, and an interesting relationship between the average number of frustrated plaquettes and the transition point at zero temperature is found.
In Sect.~3 the cases of the hierarchical lattices are examined.
We confirm the same relationship in the model with three-body interactions in Sect.~4.
The Sherrington-Kirkpatrick (SK) model is investigated in Sect.~5.
The final section is devoted to summary and discussion.

\section{Ising Spin Glasses with Binary Couplings on the Two Dimensional Lattices}

\begin{figure}[b]
  \begin{center}
  \includegraphics[width=4.5cm,clip]{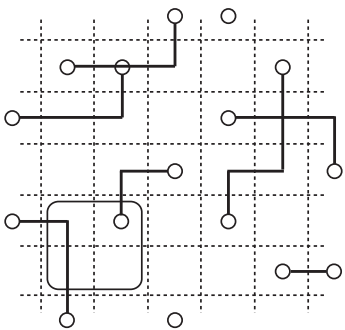}
  \caption{Antiferromagnetic bonds and frustrated plaquettes on the square lattice.
Thick segments traverse antiferromagnetic bonds, and circles denote frustrated plaquettes.
Two segments cross a loop at the lower left-hand corner.
One of the segments crosses the loop two times.
A single circle lies inside the loop.}
  \label{fig:sq_lattice}
  \end{center}
\end{figure}

Let us consider the frustration in the $\pm J$ Ising model [Eq.~(\ref{eq:pmJ})] on the square lattice.
We investigate the frustration of plaquettes, which are squares composed of four bonds in the case of the square lattice.
In Fig.~\ref{fig:sq_lattice},
dashed lines express the square lattice and antiferromagnetic bonds are traversed by thick segments.
A circle in a plaquette means that the plaquette is frustrated.
The figure shows that ends of a segment identify frustrated plaquettes.
Consider a loop on the lattice and a segment $i$.
We represent the number of intersections of the loop and the segment as $a_{i}$ and the number of ends of the segment inside the loop as $b_{i}$.
If one of the ends lies inside the loop and the other end lies outside, the segment has to cross the loop an odd number of times, that is, $a_{i}$ is an odd number.
In contrast, if both the ends lie on the same side of the loop, $a_{i}$ is an even number.
The relation
\begin{equation}
(-1)^{a_{i}} = (-1)^{b_{i}}
\end{equation}
is satisfied as a result.
Since similar equations are derived for any segment, we have 
\begin{equation}
(-1)^{\sum_{i} a_{i}} = (-1)^{\sum_{i} b_{i}},
\end{equation}
where the summation runs over segments on the lattice.
The left-hand side of this equation is the product $\prod_{<ij> \in c} (J_{ij}/J)$, where $c$ denotes the loop.
Thus, we extract only the parity of the number of circles inside the loop from the product.
When the loop contains an even number of circles, the product misses them, but no spin configuration realizes the lower-energy state in all couplings.
Hence, it is appropriate to consider elementary loops, namely plaquettes, to investigate the frustration in the system.

We calculate the configurational average over the distribution of $\{J_{ij}\}$ of the number $N_{f}^{\textrm{sq}}$ of frustrated plaquettes~\cite{Kirkpatrick, Vannimenus}.
Note that the concentration of ferromagnetic bonds is denoted by $p$ and the periodic boundary condition is assumed.
We have
\begin{equation}
\begin{split}
N_{f}^{\textrm{sq}}(p) := \big[ N_{f}^{\textrm{sq}} \big]_{p} &= \Bigg[ \sum_{\textrm{c}} \frac{1}{2} \Bigg(1 - \prod_{<ij> \in c} \frac{J_{ij}}{J} \Bigg) \Bigg]_{p} \\
&= \frac{N}{2} \{ 1 - ( 2p - 1 )^{4} \},
\end{split}
\end{equation}
where $N$ is the number of spins and $[\hspace{0.5mm} \cdot \hspace{0.5mm} ]_{p}$ denotes the configurational average with $p$.
The number of plaquettes is equal to the number of spins on the square lattice.
The summation has run over all plaquettes $c$.

We next differentiate $N_{f}^{\textrm{sq}}(p)$ with respect to the average number $N_{a}^{\textrm{sq}}(p)$ of antiferromagnet bonds, which is equal to $2N(1-p)$ in the case of the square lattice, to calculate the change in the number of frustrated plaquettes by the increase in the number of antiferromagnetic bonds.
The resulting derivative is 
\begin{equation}
v^{\textrm{sq}}(p) := \bigg[ \frac{d N_{a}^{\textrm{sq}}(p)}{d p} \bigg]^{-1} \frac{d N_{f}^{\textrm{sq}}(p)}{d p} = 2( 2p - 1 )^3. \label{eq:v}
\end{equation}
This quantity $v^{\textrm{sq}}(p)$ expresses the ratio of the increase in the number of frustrated plaquettes to that in the number of antiferromagnetic bonds.

When $p \simeq 1$, the increase in the number of frustrated plaquettes is larger than that in the number of antiferromagnetic bonds.
When $p \simeq 1/2$, the converse is realized.
There is a turning point where the increase in the number of frustrated plaquettes is equal to that in the
number of antiferromagnetic bonds.
It is identified with the point $v^{\textrm{sq}}(p^{\textrm{sq}}) = 1$, where 
\begin{equation}
p^{\textrm{sq}} =  \frac{1}{2} + \bigg(\frac{1}{2}\bigg)^{4/3} \simeq 0.8969.
\end{equation}
This value is very close to the phase transition point at zero temperature estimated by other numerical approaches (Table~\ref{table:values2D}).
\begin{table*}
\caption{Equations and solutions of $v(p) = 1$ and the locations $p_{c}$ of the transition points at zero temperature in several two-dimensional lattices.
The kagom\'{e}, dual of kagom\'{e}, extended kagom\'{e}, and dual of extended kagom\'{e} lattices are depicted in Fig.~\ref{fig:lattices}}
\label{table:values2D}
\begin{center}
\begin{ruledtabular}
\begin{tabular}{lccl}
Lattice & Our equation & Our solution & \multicolumn{1}{c}{$p_{c}$} \\
\hline
Square &  $2(2p^{\textrm{sq}}-1)^{3} = 1$ & 0.8969 & 0.8967(1)~\cite{Fujii} \\
 & & & 0.8955(11)~\cite{Jinuntuya} \\
 & & & 0.897(1)~\cite{Amoruso} \\
 & & & 0.8969(1)~\cite{Wang} \\
Triangle & $2(2p^{\textrm{tri}}-1)^{2} = 1$ & 0.8536 & 0.8412(1)~\cite{Fujii} \\
 & & & 0.833(3)~\cite{Achilles} \\
Hexagon & $2(2p^{\textrm{hex}}-1)^{5} = 1$ & 0.9353 & 0.9351(2)~\cite{Fujii} \\
 & & & 0.933~\cite{Achilles} \\
Kagom\'{e} & $(2p^{\textrm{kag}}-1)^{2} + (2p^{\textrm{kag}}-1)^{5} = 1$ & 0.9044 & 0.9052(1)~\cite{Fujii} \\
Dual of kagom\'{e} & $2(2p^{\textrm{d-kag}}-1)^{3} = 1$ & 0.8969 & 0.8837(1)~\cite{Fujii} \\
Extended kagom\'{e} & $\frac{2}{3}(2p^{\textrm{ex-kag}}-1)^{2} + \frac{4}{3} (2p^{\textrm{ex-kag}}-1)^{11} = 1$ & 0.9532 & 0.9593(2)~\cite{Fujii} \\
Dual of extended kagom\'{e} & $2(2p^{\textrm{d-ex-kag}}-1)^{2} = 1$ & 0.8536 & 0.7948(2)~\cite{Fujii} \\
\end{tabular}
\end{ruledtabular}
\end{center}
\end{table*}
This agreement suggests that the increase in the number of frustrated plaquettes controls the phases where the system lies.
Since temperature has not been considered, it is reasonable that the temperature at which our value indicates the transition point is zero.  
Moreover, our argument is consistent with the conjecture on the geometry-induced phase transitions at low temperature~\cite{Nishimori, Nishimori3}, although we have not directly observed the entropy of frustration distribution but discussed only the average number of frustrated plaquettes.

\begin{figure}
  \begin{center}
  \includegraphics[width=6.5cm,clip]{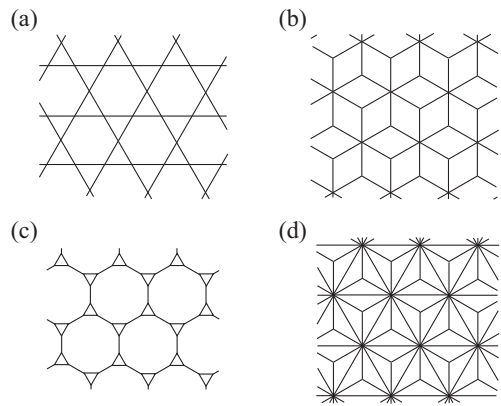}
  \caption{Two-dimensional lattices: (a) kagom\'{e}, (b) dual of kagom\'{e}, (c) extended kagom\'{e}, and (d) dual of extended kagom\'{e} lattices.}
  \label{fig:lattices}
  \end{center}
\end{figure}

Although our argument yields an extraordinarily accurate correspondence, we cannot regard it as the derivation of the phase transition point since we have not found any singularity in physical quantities.
We should examine other lattices for the reasonableness of the agreement.
The cases of other lattices are listed in Table~\ref{table:values2D}.
The solutions of $v(p) = 1$ in several lattices show fairly good agreement as in the case of the square lattice.
These results demonstrate that the correspondence between the transition point and the solution of $v(p) = 1$ is not accidental on a particular lattice.
In addition, our values in the dual of the kagom\'{e} lattice ($p^{\textrm{d-kag}} = 0.8969$) and the dual of the extended kagom\'{e} lattice ($p^{\textrm{d-ex-kag}} = 0.8536$) correspond to the values in the square lattice ($p^{\textrm{sq}} = 0.8969$) and the triangular lattice ($p^{\textrm{tri}} = 0.8536$), respectively, although the numerical estimates show differences in these lattices.
It is noted that the dual of the kagom\'{e} lattice and the dual of the extended kagom\'{e} lattice are constructed of squares and triangles, respectively, as in Fig.~\ref{fig:lattices}.
This finding shows that our argument imperfectly reflects the global structure of lattices but produces an effective approximation with the local structure of lattices.
Furthermore, it is remarkable that the result on the dual of the extended kagom\'{e} lattice has a larger deviation than the triangular lattice.
Our values on the dual of extended kagom\'{e} and triangular lattices are $p^{\textrm{d-ex-kag}} = p^{\textrm{tri}} = 0.8536$, whereas the transition points are numerically estimated as $p_{c}^{\textrm{d-ex-kag}} = 0.7948(2)$~\cite{Fujii} and $p_{c}^{\textrm{tri}} = 0.8412(1)$~\cite{Fujii}, $0.833(3)$~\cite{Achilles}, respectively.
Thus, the global structure plays an essential role in the phase transition in the model on the former lattice.
Indeed, we can find other geometries, for example hexagons, in the dual of extended kagom\'{e} lattice.

\section{Hierarchical Lattices}

\begin{figure}
  \begin{center}
  \includegraphics[width=4.5cm,clip]{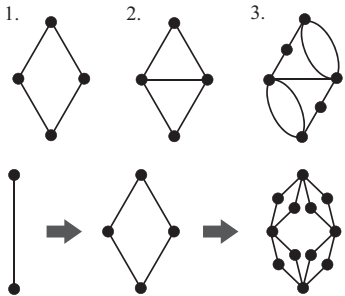}
  \caption{Units of hierarchical lattices~1, 2, and 3 examined in the present paper and the replacement of each single bond with the unit of lattice~1.}
  \label{fig:hierarchical}
  \end{center}
\end{figure}

We next apply the method developed in the last section to the $\pm J$ Ising model on the hierarchical lattices~\cite{Berker}.
The lattices are generated by iterating the process that each single bond is replaced with a unit of the lattice, as depicted in Fig.~\ref{fig:hierarchical}.
We can obtain the accurate transition point on the hierarchical lattices with renormalization-group calculations~\cite{Berker, Nobre}.
The lattices are, thus, a good testing ground for the comparison of estimates of the transition point.
In addition, it is valuable to examine whether or not our argument yields the accurate transition points in unusual lattices.
We examine lattices~1, 2, and 3 in Fig.~\ref{fig:hierarchical}.

Let us calculate the average number of frustrated plaquettes in the hierarchical lattices.
We first consider lattice 1.
There are plaquettes of various shapes in the lattice, but a rule is found.
The plaquette of the type generated for the first time by the $k$-th substitution is composed of $2^{k+1}$ bonds.
For example, a plaquette composed of $8 \ (=2^{2+1})$ bonds appears in the second substitution (Fig.~\ref{fig:hierarchical}).
The number of plaquettes of $2^{k+1}$ bonds in the lattice produced by iterating the replacement process $n$ times is $4^{n-k}$, since the number of plaquettes of a particular type quadruples in a single replacement. 
For example, the number of plaquettes of $4 \ (=2^{1+1})$ bonds in the lattice after the second substitution is $4 \ (=4^{2-1})$ (Fig.~\ref{fig:hierarchical}).
With the periodic boundary condition, we have another plaquette composed of $2^{n+1}$ bonds on the surface of the lattice.
The average number of frustrated plaquettes is, therefore, represented as
\begin{equation}
N_{f}^{\textrm{hier-1}}(p) = \sum_{k=1}^{n} 4^{n-k} 2^{\delta_{n,k}} \frac{1}{2} \Big\{ 1 - ( 2p - 1 )^{2^{k+1}} \Big\}.
\end{equation}
The derivative corresponding to Eq.~(\ref{eq:v}) in the infinite-volume limit is
\begin{equation}
v^{\textrm{hier-1}}(p) = \lim_{n \to \infty} \sum_{k=1}^{n} 2^{-k+1} (2p-1)^{2^{k+1} -1}.
\end{equation}
Note that the average number $N_{a}^{\textrm{hier-1}}(p)$ of antiferromagnetic bonds is $4^{n}(1-p)$.
Evaluating the solution of $v^{\textrm{hier-1}}(p^{\textrm{hier-1}}) = 1$ in the case of $n = 20$, we obtain $p^{\textrm{hier-1}} = 0.9477$.
Since the same value is obtained even in the case of $n = 30$, it is reasonable to interpret this value as the actual solution.

In order to compare this value with the location of the transition point at zero temperature, we derive the transition point by the renormalization-group method proposed by Nobre~\cite{Nobre}.
We first prepare a pool of $M$ couplings following the $\pm J$ distribution, where we set $M$ and $J$ equal to $10^{6}$ and $1$, respectively.
A unit of the lattice is built of bonds randomly taken from the pool.
The unit is renormalized in the ordinary manner of the real-space renormalization group, which is the reverse operation of the construction of the lattice.
Then, a renormalized bond is produced and we put it in a new pool.
This scheme is executed $M$ times and the new pool is filled by $M$ renormalized bonds as a result.
This is a single renormalization-group transformation for the whole system.
After the renormalization for the whole system $30$ times, we determine the transition point in terms of the average of the resulting couplings.
If the average is larger (smaller) than $10$ ($10^{-3}$), we regard the system to be in the ferromagnetic (paramagnetic or spin-glass) phase.
To reduce statistical errors, we run $100$ samples.

The solutions of $v(p) = 1$ have a small difference but are qualitatively close to the estimates of the transition points at zero temperature in the respective lattices listed in Table~\ref{table:valueshier}.
\begin{table*}
\caption{The equations and solutions of $v(p) = 1$ and the locations $p_{c}$ of the transition points at zero temperature in the hierarchical lattices in Fig.~\ref{fig:hierarchical}.
The indices $k$ of summations in the equations run from $1$ to infinity.}
\label{table:valueshier}
\begin{center}
\begin{ruledtabular}
\begin{tabular}{cccl}
Lattice & Our equation & Our solution & \multicolumn{1}{c}{$p_{c}$} \\
\hline
1 &  $\sum 2^{-k+1} (2p^{\textrm{hier-1}}-1)^{2^{k+1} -1} = 1$ & 0.9477 & 0.9215(3) \\
2 & $\sum 3 \left( \frac{2}{5} \right)^{k} (2p-1)^{3 \cdot 2^{k-1} -1} = 1$ & 0.9089 & 0.8935(1) \\
3 & $\sum 3^{-k-1} \big\{ 4 (2p^{\textrm{hier-3}}-1)^{2 \cdot 3^{k-1} -1} + 8 (2p^{\textrm{hier-3}}-1)^{4 \cdot 3^{k-1} -1} \big\} = 1$ & 0.9187 & 0.8951(3) \\
\end{tabular}
\end{ruledtabular}
\end{center}
\end{table*}
It is remarkable that the results correctly express the order of locations of the transition points.
In particular, a small difference between lattices 2 and 3 is distinguished.
This result also demonstrates the effectiveness of our argument to qualitatively predict the transition points on the unusual lattices.

\section{Three-Body Ising Spin Glasses with Binary Couplings}

In this section, we apply our method to a different type of model from previous ones, the $\pm J$ Ising model with the three-body interactions,
\begin{equation}
H = -\sum_{<i,j,k>} J_{ijk} S_{i} S_{j} S_{k}.
\end{equation}
Couplings $J_{ijk}$ are governed by the same distribution as the previous models with the two-body interactions.
One of the reasons for investigating this model is that the order-disorder phase boundary in this model is equivalent to the accuracy threshold in a type of topological quantum error-correcting code, namely, color codes~\cite{Bombin, Katzgraber}.

\begin{figure}
  \begin{center}
  \includegraphics[width=3cm,clip]{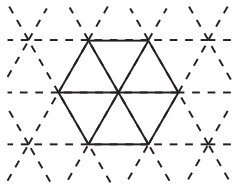}
  \caption{Unit of frustration in the $\pm J$ model with the three-body interactions on the triangular lattice.
The unit is a hexagon composed of six triangles depicted by thick lines.}
  \label{fig:three-body}
  \end{center}
\end{figure}

Let us consider the model on the triangular lattice to illustrate the frustration in this model.
Three spins on the same triangle interact with each other and couplings $J_{ijk}$ reside on triangles.
A loop in this model is composed of triangles.
Each triangle in a loop connects to the neighboring triangles on one or two vertices.
We call the minimal loop, which can be frustrated, the ``unit of frustration".
The unit of frustration in the triangular lattice is a hexagon composed of six triangles depicted as Fig.~\ref{fig:three-body}.
If the number of negative couplings in a unit of frustration is odd, the unit is frustrated.
However, this simple rule does not determine whether an arbitrary loop other than the unit of frustration is frustrated or not.
Even if the number of negative couplings on the loop is odd, it is possible that a spin configuration realizes the lower-energy state in all couplings.
On the other hand, whenever there is no frustrated unit in a loop, we can find the spin configuration that satisfies all couplings.
In other words, loops are frustrated only with frustrated units.
This finding means that units of frustration can be used to study the nature of frustration in the system, and larger loops can miss details of frustration as in the previous models with two-body interactions. 

The average number $N_{f}^{\textrm{3,tri}}(p)$ of frustrated units is represented as 
\begin{equation}
N_{f}^{\textrm{3,tri}}(p) = \frac{N}{2} \{ 1 - (2p-1)^{6} \},
\end{equation}
where $N$ denotes the number of spins equal to the number of units of frustration in this lattice.
The average number $N_{a}^{\textrm{3,tri}}$ of negative triangles is $2N(1-p)$, and the derivative of $N_{f}^{\textrm{3,tri}}(p)$ with respect to $N_{a}^{\textrm{3,tri}}(p)$ is
\begin{equation}
v^{\textrm{3,tri}} (p) = 3 (2p-1)^{5}.
\end{equation}

We obtain $p^{\textrm{3,tri}} = 0.9014$, where $v^{\textrm{3,tri}} (p^{\textrm{3,tri}}) = 1$.
Since the transition point at zero temperature in the model, to the best of our knowledge, has not been numerically estimated in any other studies, we compare this value with the concentration of negative triangles at the transition point on the NL instead of the point at zero temperature.
The critical concentrations on the NL and at zero temperature are expected to be very similar.
Indeed, the two values are very close in the case of the Union-Jack lattice, where the transition point at zero temperature has been estimated~\cite{Landahl}.
The solutions of $v(p) = 1$ are close to the critical concentrations at zero temperature or on the NL (Table~\ref{table:values3}).
\begin{table*}
\caption{Equations and solutions of $v(p) = 1$ and the critical concentrations $p_{c}$ at zero temperature and $p_{m}$ on the NL in the three-body $\pm J$ model on the triangular and Union-Jack lattices.
The critical concentration at zero temperature is expected to be slightly smaller than that on the NL in the triangular lattice as in the Union-Jack lattice.}
\label{table:values3}
\begin{center}
\begin{ruledtabular}
\begin{tabular}{lccll}
Lattice & Our equation & Our solution & \multicolumn{1}{c}{$p_{c}$} & \multicolumn{1}{c}{$p_{m}$} \\
\hline
Triangle &  $3(2p^{\textrm{3,tri}}-1)^{5} = 1$ & 0.9014 & & 0.8903(1)~\cite{Ohzeki3} \\
 & & & & 0.891(2)~\cite{Katzgraber} \\
Union-Jack & $(2p^{\textrm{3,U-J}}-1)^{3} + 2(2p^{\textrm{3,U-J}}-1)^{7} = 1$ & 0.9059 & 0.8944(1)~\cite{Landahl} & 0.89075(5)~\cite{Ohzeki3} \\
 & & & & 0.891(2)~\cite{Katzgraber2} \\
\end{tabular}
\end{ruledtabular}
\end{center}
\end{table*}

Comparing the result for the triangular lattice with the results for the models with the two-body interactions on the triangular and hexagonal lattices shown in Table~\ref{table:values2D}, we find that our argument properly distinguishes the number of bodies in the couplings.
In particular, the difference between the three-body $\pm J$ model on the triangular lattice and the two-body $\pm J$ model on the hexagonal lattice is worthy of attention, since the units of frustration in these two models have the same shape.
This coincidence leads to the same term $(2p-1)^{5}$ in both functions $v(p)$ accordingly.
The coefficients of the terms, $3$ in the three-body $\pm J$ model and $2$ in the two-body $\pm J$ model, generate an appropriate difference in the results.
The coefficients express the increment of frustrated units by an additional negative coupling to the perfect pure system.

\section{SK Model}

We investigate the standard infinite-range model in spin glasses called the SK model~\cite{Sherrington},
\begin{equation}
H = - \sum_{i<j} J_{ij} S_{i} S_{j}, \label{eq:SK}
\end{equation}
where any two spins interact with each other.
The couplings $J_{ij}$ are governed by the Gaussian distribution, 
\begin{equation}
P(J_{ij}) = \sqrt{\frac{N}{2 \pi}} \exp \bigg[ - \frac{N}{2} \left(J_{ij} - \frac{J_{0}}{N} \right)^{2} \bigg], \label{eq:gauss}
\end{equation}
where $N$ denotes the number of spins, and $N$ is introduced in the distribution to make the resulting extensive quantities proportional to $N$.

We consider the number of frustrated plaquettes.
A plaquette in this model is a triangle composed of three bonds.
Since quenched variables $J_{ij}$ can take arbitrary real values, their product over a plaquette also takes arbitrary values.
However, we do not take account of the magnitude of the product.
Our attention is focused on whether or not a plaquette is frustrated.
This is one of the methods of generalizing the method developed in previous cases.
The average number of frustrated plaquettes is represented as 
\begin{equation}
\begin{split}
N_{f}^{\textrm{SK}}(J_{0}) &= \Bigg[ \sum_{c} \ \frac{1}{2} \Bigg( 1 - \prod_{<ij> \in c} \frac{J_{ij}}{|J_{ij}|} \Bigg) \Bigg]_{J_{0}} \\
&= \frac{N(N-1)(N-2)}{6} \\
& \hspace{1cm} \times \frac{1}{2} \bigg\{ 1 - \bigg[ \textrm{erf} \left( \frac{J_{0}}{\sqrt{2N}} \right) \bigg]^{3} \bigg\},
\end{split}
\end{equation}
where $N(N-1)(N-2)/6$ expresses the number of plaquettes equal to the number of combinations of three arbitrary sites on the lattice, and $\textrm{erf} (\cdot)$ is the error function
\begin{equation}
\frac{1}{2} \textrm{erf} \left( \frac{x}{\sqrt{2}} \right) = \int_{0}^{x} dz \frac{1}{\sqrt{2 \pi}} \exp \bigg[ - \frac{(z - x)^{2}}{2} \bigg].
\end{equation}
The average number of antiferromagnetic bonds is
\begin{equation}
N_{a}^{\textrm{SK}}(J_{0}) = \frac{N(N-1)}{2} \ \frac{1}{2} \bigg[ 1 - \textrm{erf} \left( \frac{J_{0}}{\sqrt{2N}} \right) \bigg].
\end{equation}
In the infinite-volume limit $N \to \infty$, the derivative of $N_{f}^{\textrm{SK}}(J_{0})$ with respect to $N_{a}^{\textrm{SK}}(J_{0})$ is
\begin{equation}
v^{\textrm{SK}}(J_{0}) := \lim_{N \to \infty} \bigg[ \frac{d N_{a}^{\textrm{SK}}(J_{0})}{d J_{0}} \bigg]^{-1} \frac{d N_{f}^{\textrm{SK}}(J_{0})}{d J_{0}}
= \frac{2}{\pi} J_{0}^{2}.
\end{equation}
The solution of the equation $v^{\textrm{SK}}(J_{0}^{\textrm{SK}}) = 1$ is $J_{0}^{\textrm{SK}} = \sqrt{\pi/2}$.
This result is not in agreement with the well-known exact transition point at zero temperature $J_{0} = 1$~\cite{Toulouse2}, but is identical to the transition point under the assumption of replica symmetry~\cite{Sherrington}, which is derived from the equation of state of the ferromagnetic order parameter $m$ at zero temperature:
\begin{equation}
m = \textrm{erf} \left( \frac{J_{0}}{\sqrt{2}} m \right). 
\label{eq:stateofeq_sk}
\end{equation}
This equation has a solution with $m \neq 0$ if the slope of the function on the right-hand side at $m=0$, which is $\sqrt{2/\pi} J_{0}$, is larger than $1$.
Therefore, the phase transition point is determined by $v^{\textrm{SK}}(J_{0}) = 1$, which is identical to the condition in our argument.
Our result with frustration accords with the replica-symmetric solution not only in the solution but also in the equation for the determination of the transition point.

\section{Summary and Discussion}

We have reported the relationship between frustration and the transition point of Ising spin glasses.
The relation reveals that the concentration of ferromagnetic bonds in the system is rather close to the critical concentration at zero temperature when the derivative of the average number of frustrated plaquettes with respect to the average number of antiferromagnetic bonds is equal to unity.
This relation is confirmed in various models.
In particular, the cases in the two-dimensional lattices show good correspondence.
In the SK model, the solution derived from our argument is in exact agreement with the transition point under the assumption of replica symmetry, and the slope of the average number of frustrated plaquettes corresponds to the slope of the function in the equation of state of the ferromagnetic order parameter at zero temperature.

Although we have detected no conventional sign of phase transitions, the agreements between our values and the transition points are naturally regarded not to be accidental.
One of the reasons is that our argument attains pretty accurate predictions in most of the models that we have applied the method to.
The variety of models in which the agreements are found is particularly remarkable. 
Our argument is, thus, expected to yield approximate locations of the transition points.

Our approach to spin glasses by means of frustration is novel and different from conventional ones.
The representative conventional study is an attempt to characterize phases in terms of the change in the free energy by increasing frustration~\cite{Fradkin}.
Both our study and the conventional one focus on the change of frustration to discuss phases.
Hence, they might have a deep connection, but we have not found it yet.
It is notable that our approach enables us to extract the approximate locations of the transition points specifically, which is difficult for conventional ones.

Furthermore, it is remarkable that the result for the SK model, which corresponds not to the exact solution but to the replica symmetric one, shows deep significance of the deviation from the exact solution.
This result suggests that we should not disregard the gap between our solution of $v = 1$ and the exact one as a simple defect of our method.
Moreover, the correspondence of $v^{\textrm{SK}}(J_{0})$ and the slope of the equation of state provides an insight into our heuristic condition $v = 1$.
Also, in other cases, $v$ is expected to be concerned with the slope of the equation of state under some assumption such as that of replica symmetry.

The simplicity of our calculation in contrast with the accuracy of the results has practical use in the context of topological quantum error-correcting codes, where the accuracy threshold corresponds to the phase boundary of spin glasses~\cite{Dennis}.
Our method is useful for the first approximation of the accuracy threshold.
The simplicity, moreover, illuminates the unknown elegant physics in spin glasses, which is usually regarded as a complex subject.

Since we have considered only frustration to discuss phase transitions, our result seems to support the conjecture on the geometry-induced phase transition~\cite{Nishimori, Nishimori3}.
Nevertheless, our values are different from the expected locations from the conjecture~\cite{Nishimori, Nishimori3, Ohzeki, Ohzeki2, Hasenbusch, Queiroz, Katzgraber, Ohzeki3, Katzgraber2}.
The reason is that we have observed the number of frustrated plaquettes, whereas the entropy of frustration distribution has been considered in the conjecture~\cite{Nishimori, Nishimori3}.
Our result suggests the existence of the geometry-induced phase transition, but it does not support the proposition that the root of the transition at zero temperature is a singularity in the entropy of frustration distribution.

We should mention higher-dimensional cases.
Our method is not limited to the two-dimensional lattices and can be applied to higher-dimensional ones since only the local structure of a lattice, namely bonds and plaquettes, is needed to execute our scheme.
In the three-dimensional cubic lattice, for example, a plaquette consists of four bonds as in the square lattice, and we have a function $v^{\mathrm{cube}}(p) = 4 (2p-1)^{3}$ for the $\pm J$ Ising model with the two-body interactions.
The condition $v^{\mathrm{cube}}(p^{\mathrm{cube}}) = 1$ yields the value $p^{\mathrm{cube}} = 0.8150$, while a numerical study~\cite{Thomas} estimates the transition point at zero temperature to be $p_{c}^{\mathrm{cube}} = 0.7747(7)$.
The two values are not so close, but our calculation gives a fairly good approximate value despite its simplicity.
This result demonstrates the usefulness of our method for the higher-dimensional cases, where duality transformation, one of the standard tools to derive the locations of transition points, cannot be used basically.

Since our results suggest the relationship among frustration, phases, and the replica symmetry, the origin of the correspondence between our results from frustration and the transition points at zero temperature might lead to clarifying the fruitful structure behind spin glasses in general.
The origin is, however, not obvious yet.
To clarify it, further investigation is necessary.
In particular, the lack of an effect from the global structure of lattices and from the nature of spins should be resolved.
It is particularly important to clarify the reason why our results are consistent with the transition points of models with the Ising spins.
In addition, the validity of the naive generalization of the method to the higher-dimensional systems including the SK model should be further studied.

\vspace{0.7cm}
\begin{acknowledgments}
The author is grateful to H. Nishimori, M. Ohzeki, K. Takahashi, and K. Takeda for valuable discussions and thanks K. Fujii and Y. Tokunaga for providing the unpublished data in Table~\ref{table:values2D}.
This work was partially supported by a Research Fellowship from the Japan Society for the Promotion of Science.
\end{acknowledgments}


\end{document}